\documentstyle[aps,prb,multicol]{revtex}
\input epsf
\newcommand{\vf}{{\bf v}_f}

\renewcommand{\vec}[1]{{\bf{#1}}}

\newcommand{\beq} {\begin{equation}}
\newcommand{\eeq} {\end{equation}}
\newcommand{\ber} {\begin{eqnarray}}
\newcommand{\eer} {\end{eqnarray}}

\newcommand{\de}   {{\delta}}

\newcommand{\ts}{\textstyle }

\newcommand{\lsim}{ {{}_{\ts <} \atop {\ts \sim}} }
\newcommand{\gsim}{ {{}_{\ts >} \atop {\ts \sim}} }

\newcommand{\ep}{\epsilon }

\newcommand{\mat}{\left(\begin{array}{cc}}
\newcommand{\matend}{\end{array}\right)}
\newcommand{\matl}{\left(\begin{array}{c}}
\newcommand{\matlend}{\end{array}\right)}
\newcommand{\qt}{{\, \scriptstyle \otimes}\, }

\newcommand{{\ra}}{{\scriptscriptstyle R,A}}
\newcommand{{\rak}}{{\scriptscriptstyle R,A,K}}
\newcommand{{\ret}}{{\scriptscriptstyle R}}
\newcommand{{\adv}}{{\scriptscriptstyle A}}
\newcommand{\kel}{{\scriptscriptstyle K}}

\newcommand{\ga}{\gamma}
\newcommand{\gb}{\tilde{\ga}}
\newcommand{\gar}{\ga^{\ret}}
\newcommand{\gbr}{\gb^{\ret}}
\newcommand{\gaa}{\ga^{\adv}}
\newcommand{\gba}{\gb^{\adv}}
\newcommand{\gara}{\ga^{\ra}}
\newcommand{\gbra}{\gb^{\ra}}

\newcommand{\xa}{x^{\kel}}
\newcommand{\xb}{\tilde{x}^{\kel}}

\newcommand{\Da}{\Delta}

\newcommand{\Db}{\tilde{\Da}}
\newcommand{\Dar}{\Da^{\ret}}
\newcommand{\Dbr}{\Db^{\ret}}
\newcommand{\Daa}{\Da^{\adv}}
\newcommand{\Dba}{\Db^{\adv}}
\newcommand{\Dara}{\Da^{\ra}}
\newcommand{\Dbra}{\Db^{\ra}}

\newcommand{\Dak}{\Da^{\kel}}
\newcommand{\Dbk}{\Db^{\kel}}

\newcommand{\va}{v}
\newcommand{\vb}{\tilde{\va}}
\newcommand{\vak}{\va^{\kel}}
\newcommand{\vbk}{\vb^{\kel}}
\newcommand{\vara}{\va^{\ra}}
\newcommand{\vbra}{\vb^{\ra}}
\newcommand{\var}{\va^{\ret}}
\newcommand{\vbr}{\vb^{\ret}}
\newcommand{\vaa}{\va^{\adv}}
\newcommand{\vba}{\vb^{\adv}}

\newcommand{\qpartial}{\vf \cdot \mbox{\boldmath $\nabla $}}
\newcommand{\grad}{\mbox{\boldmath $\nabla $}}

\begin{document}

\title{Electromagnetic Response of a Vortex in Layered Superconductors}
\author{Matthias Eschrig, J. A. Sauls}
\address{Department of Physics \& Astronomy,
         Northwestern University, Evanston, IL 60208, USA}
\author{D. Rainer}
\address{Physikalisches Institut, Universit\"at Bayreuth, D-95440 Bayreuth, Germany}
\maketitle
\begin{abstract} 
We calculate the response of a vortex core in a
layered superconductor to {\em ac} electromagnetic 
fields with frequencies $\omega\lsim 2\Delta/\hbar$. 
In this frequency range the response is 
dominated by order para\-meter collective modes
which are coupled to the vortex-core bound states of
Caroli, de Gennes and Matricon. 
Our calculations show that a vortex core 
has  a more complex and richer dynamics
than predicted by previous theories.
The {\em ac} field drives an oscillating, nearly homogeneous 
supercurrent in the direction of the electric field,
superimposed with a dissipative current flow which has
a dipolar spatial structure.
The order parameter response at low frequencies is 
an approximately rigid collective motion of the vortex
structure perpendicular to the external field. This
structure becomes strongly deformed
at frequencies of order $\omega\gsim 0.5 \Delta /\hbar $.
Coupling of the vortex-core
bound states to collective modes of the order parameter
at low frequencies leads to substantial enhancement of
the dissipation near the vortex-center well above that of the
normal-state.
\end{abstract}
\medskip

\begin{multicols}{2}
Vortex cores play a key role in dissipation processes 
of superconductors in the
Abrikosov phase. Bardeen and Stephen\cite{Bardeen65}
modeled the vortex core as a region of normal metal
in which the excitations in the core respond to an electromagnetic
field like electrons in the normal metallic state.
This is a good approximation for dirty superconductors
with a mean free path, $\ell$, much smaller than the
coherence length $\xi_0$. However, in clean superconductors
the low-lying excitations in the
core are the bound states of Caroli, 
de Gennes and Matricon.\cite{Caroli64}
These excitations have superconducting and normal properties.
They are the source of circulating supercurrents in the
equilibrium vortex core,\cite{Bardeen69}
and they are strongly coupled to the condensate by
Andreev processes.\cite{Bardeen69,Rainer96} Their response
to an electromagnetic field is radically different from 
that of normal electrons.
For vortex cores one has two fundamentally different 
origins of dissipation. One is dissipation by the collective
motion of the whole condensate, and the second is dissipation
by transitions between Caroli-de Gennes-Matricon
bound states. These processes are coupled because of the
strong interaction between the condensate and the bound states, 
and requires a self-consistent treatment of condensate and bound-state
dynamics. Earlier calculations 
neglected this coupling.\cite{Janko92,Kopnin98}
We  present the first fully self-consistent
calculation of the current response 
to an {\em ac} electric field, with frequencies comparable with the
gap frequency $\Delta/\hbar$,
in the vortex core.
Our results show that self-consistency is essential for even a qualitative
understanding of low-frequency dynamics and dissipation by the core.

We consider an s-wave superconductor 
with a random distribution
of impurities in a static magnetic field.
The applied {\it ac} electric field,
$\delta\vec{E}(t)=-\frac{1}{c} \partial_t \delta \vec{A}(t)$, 
is linearly polarized in $\hat{\vec x}$-direction,
and its wavelength is large compared
to $\xi_0$.
We assume an impurity scattering rate such
that the superconductor is outside the
superclean limit; i.e.
all bound states are broadened by an amount
comparable to or larger than
the ``mini-gap'', $\Delta^2/E_f$.\cite{Caroli64}
We investigate the intermediate clean regime,
$\xi_0\lsim\ell\lsim(E_f/\Delta)\xi_0$,
where we expect the model of a ``normal metal core''
to break down. In the intermediate clean range we can
use the quasiclassical theory of 
Fermi liquid superconductivity formulated by Eilenberger,
\cite{Eilenberger68} Larkin and Ovchinnikov,\cite{Larkin68}
and Eliashberg,\cite{Eliashberg72} which is a powerful method for
studying non-equilibrium superconductivity.

The numerical solution of nonequilibrium transport
problems is greatly simplified by a new formulation of
the nonequilibrium quasiclassical equations.
Two of the new
transport equations are a generalization of the 
Riccati-type equilibrium equations of 
Nagato et al.,\cite{Nagato93} and Schopohl and Maki.\cite{Schopohl95}
We present the central equations of the theory and 
refer readers to Ref.\onlinecite{Eschrig97} for notation, their derivation 
and the connection to the quasiclassical Green's functions.

The scalar transport equations for the distribution functions,
$\gara$, $\gbra$, $\xa$, $\xb$, which
are functions of momentum ${\bf p}_f$, position
${\bf R}$, energy $\epsilon$, and time $t$, are
\ber
\label{cricc1}
&&\displaystyle
i\hbar \, \qpartial \gara +2\ep \gara =
-\gara \qt \Dbra \qt \gara +  \nonumber \\
&& \displaystyle
\qquad \qquad \vara \qt \gara - \gara \qt \vbra - \Dara \\
\label{cricc2}
&&\displaystyle
i\hbar \, \qpartial \gbra -2\ep \gbra =
-\gbra \qt \Dara \qt \gbra + \nonumber \\
&& \displaystyle
\qquad \qquad \vbra \qt \gbra - \gbra \qt \vara  - \Dbra  \\
\label{keld1}
\displaystyle
&&i\hbar \, \qpartial \xa + i\hbar \, \partial_t \xa + \nonumber \\
&& \displaystyle
\Big(\gar \qt \Dbr -\var\Big) \qt \xa +
\xa \qt \Big( \Daa \qt \gba +\vaa \Big) = \nonumber \\
&& \displaystyle
\qquad  \qquad \gar \qt \vbk \qt \gba
-\Dak \qt \gba - \gar \qt \Dbk -\vak \\
\label{keld2}
&&\displaystyle
i\hbar \, \qpartial \xb - i\hbar \, \partial_t \xb + \nonumber \\
&& \displaystyle
\Big( \gbr \qt \Dar -\vbr\Big) \qt \xb +
\xb \qt \Big( \Dba \qt \gaa + \vba \Big)= \nonumber \\
&& \displaystyle
\qquad  \qquad \gbr \qt \vak \qt \gaa
-\Dbk \qt \gaa - \gbr \qt \Dak - \vbk .
\eer
External fields, the superconducting order parameter and impurity scattering
enter the transport equations as driving fields and scattering integrals
through the self-energies,
$v^{R,A,K}$, $\tilde v^{R,A,K}$, $\Delta^{R,A,K}$, and $\tilde\Delta^{R,A,K}$.
These self energies are functionals
of the distribution functions and must be determined self-consistently.
Equations (\ref{cricc1}-\ref{keld2}), together with the self-energy equations,
are numerically stable and are the basis for efficient 
numerical algorithms for solving nonequilibrium problems in the quasiclassical
theory of superconductivity. For the {\it ac} response of the vortex
we linearize Eqs. (\ref{cricc1}-\ref{keld2}) 
in the external field.\cite{Eschrig97}
The linear response equations require as 
input the self-consistent equilibrium 
functions, $\gamma^{R,A}_{\mbox{\small eq}}$ and 
$\tilde{\gamma}^{R,A}_{\mbox{\small eq}}$, the
order parameter and the impurity $t$-matrix for a pancake 
vortex. With these equilibrium solutions we then solve
the first-order transport equations together
with the linearized self-energy equations and the 
charge neutrality condition.
The linearization is justified as long as the relation
$e|\delta \vec{E}| \ll (\hbar \omega )^2/\Delta \xi_0 $ holds. \\
The results presented below are calculated for a
layered s-wave superconductor with a cylindrical Fermi
surface along the $\hat{\vec c}$-direction,
isotropic Fermi velocity ${\bf v}_f$
in the ab-plane, isotropic pairing interaction and a 
large Ginzburg-Landau parameter, $\kappa\gg 1$.
Impurity scattering is taken into account self-consistently
in the Born approximation. 
We consider a moderately clean superconductor with
a long  mean free path,
$\ell=10\xi_0$, where $\xi_0=\hbar v_f/2\pi k_B T_c$
is the coherence length,
and choose a  low temperature, $T=0.3T_c$.\\
\begin{figure}[tb]
\centerline{\epsfxsize0.90\hsize\epsffile{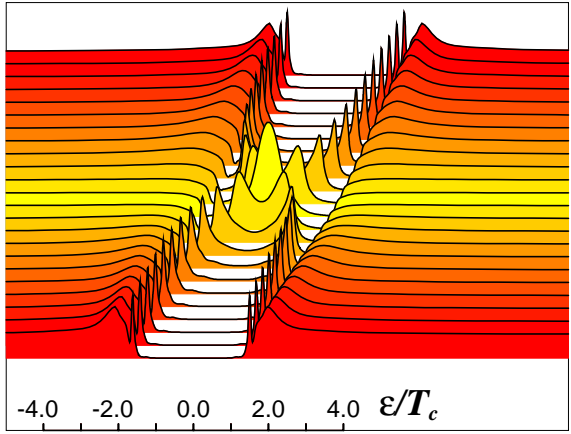}}
\vspace{3mm}
\begin{minipage}{0.95\hsize}
\caption[]{\label{Nofeps}\small
Local density of states at different distances (up
to $8.8\xi_0$, spacing $0.8\xi_0$)
from the vortex center. The center of the vortex corresponds to the 
bright filled spectrum.
}
\end{minipage}
\end{figure}
Fig. \ref{Nofeps} shows the calculated local density of states
of an  equilibrium vortex at various distances 
from the core center. The important feature 
at the vortex center is the zero-energy
bound state, which is 
broadened by impurity scattering into a resonance
of width $\Gamma \approx 0.6 T_c$. 
At finite distances from the vortex center the bound states
corresponding to different impact parameters of quasiclassical
trajectories form a one-dimensional band, which is broadened by 
impurity scattering.
The bands widen because of the coupling to the vortex flow
field with increasing distance from the center,
and develop Van Hove singularities at the band edges.
\begin{figure}[tb]
\noindent
\begin{minipage}[t]{4.0cm}
\centerline{
\epsfxsize3.9cm
\epsffile{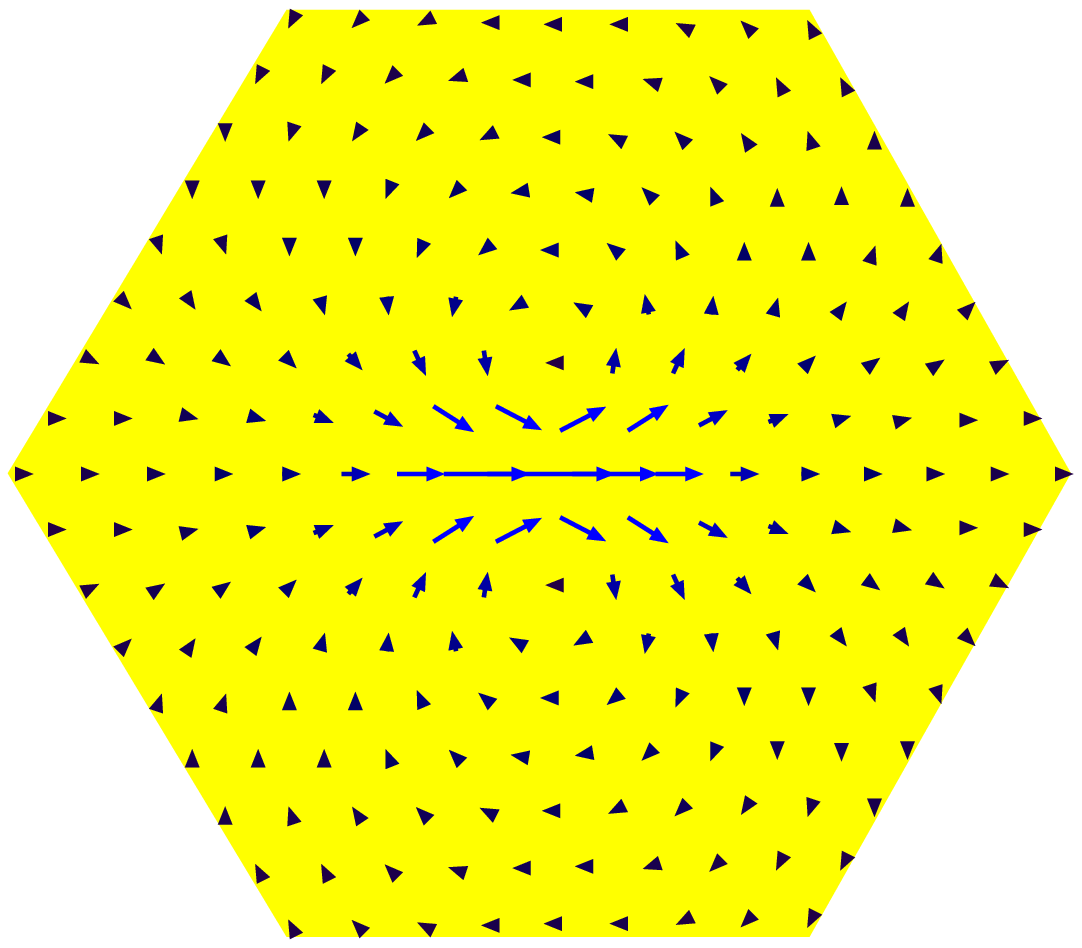}
}
\end{minipage}
\begin{minipage}[t]{4.0cm}
\centerline{
\epsfxsize3.9cm
\epsffile{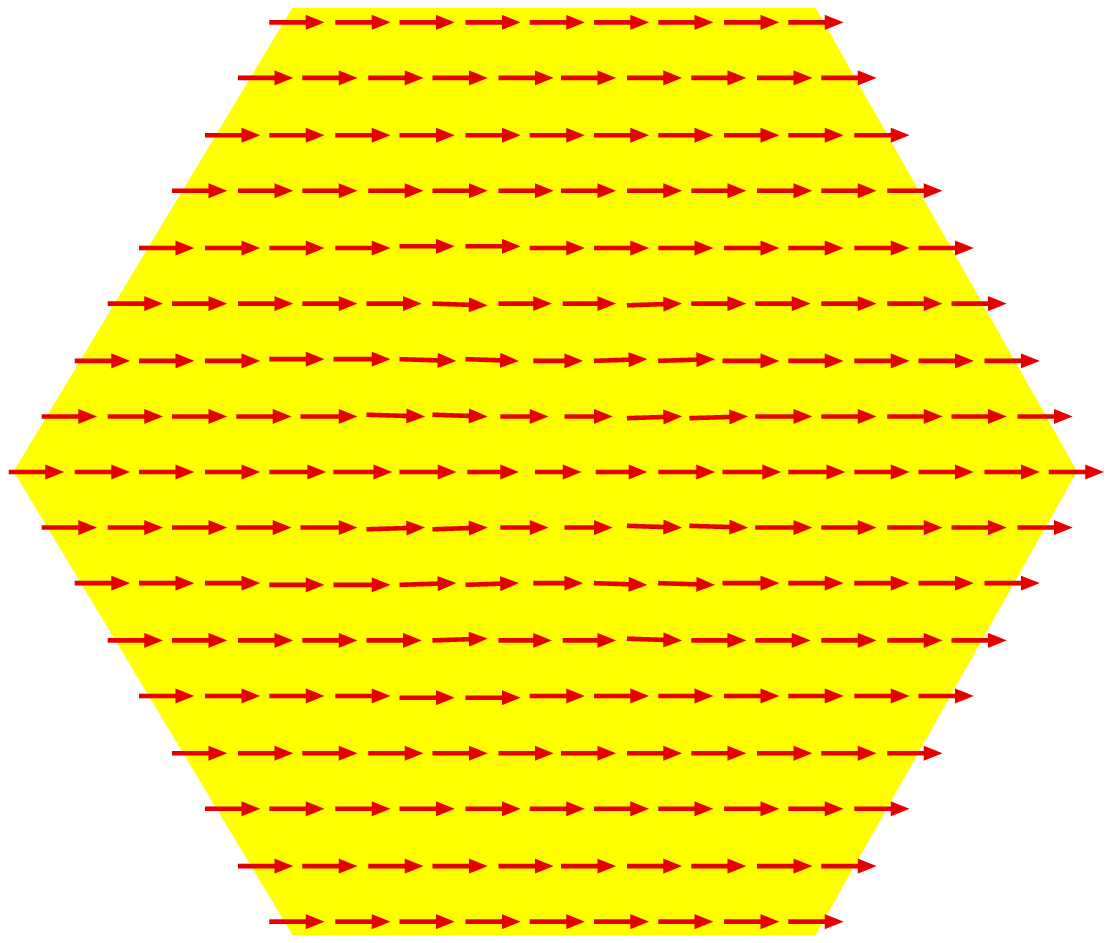}
}
\end{minipage}\\[3mm]
\begin{minipage}[t]{4.0cm}
\centerline{
\epsfxsize3.9cm
\epsffile{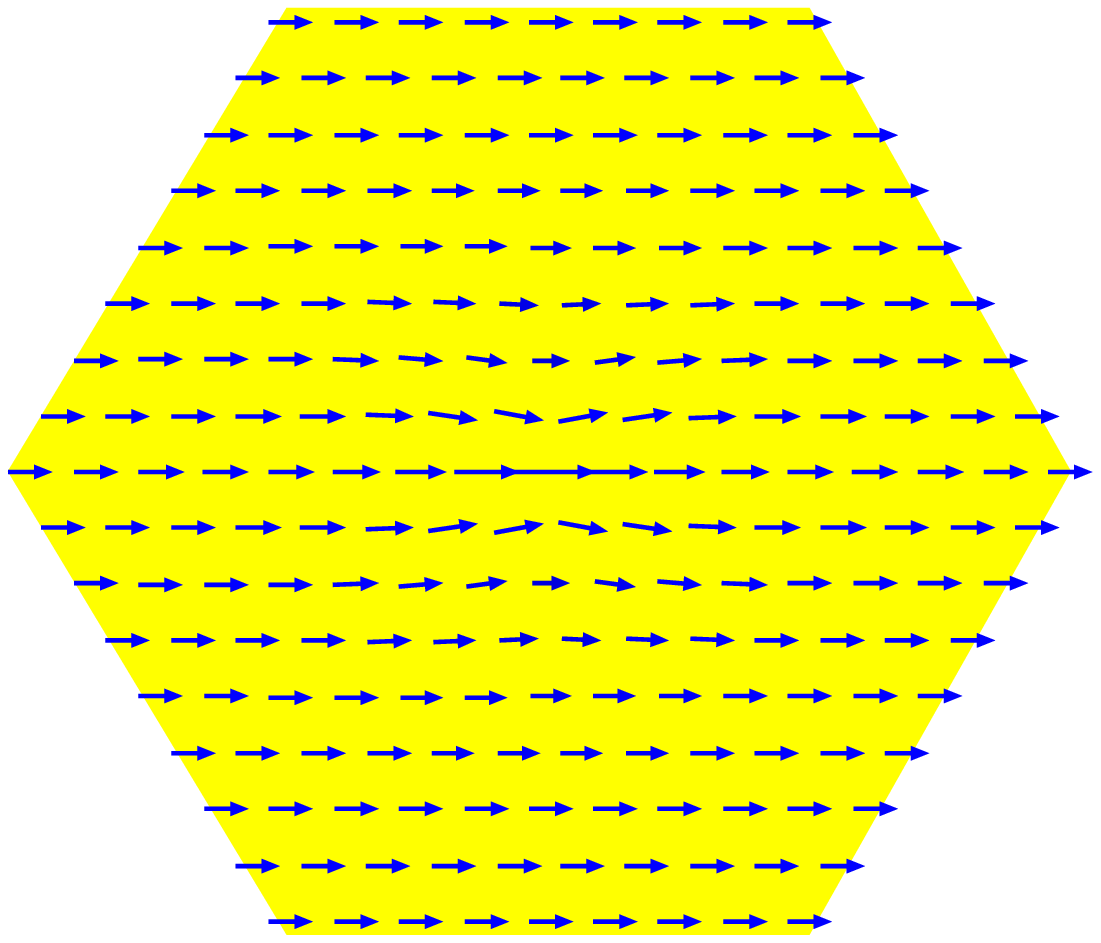}
}
\end{minipage}
\begin{minipage}[t]{4.0cm}
\centerline{
\epsfxsize3.9cm
\epsffile{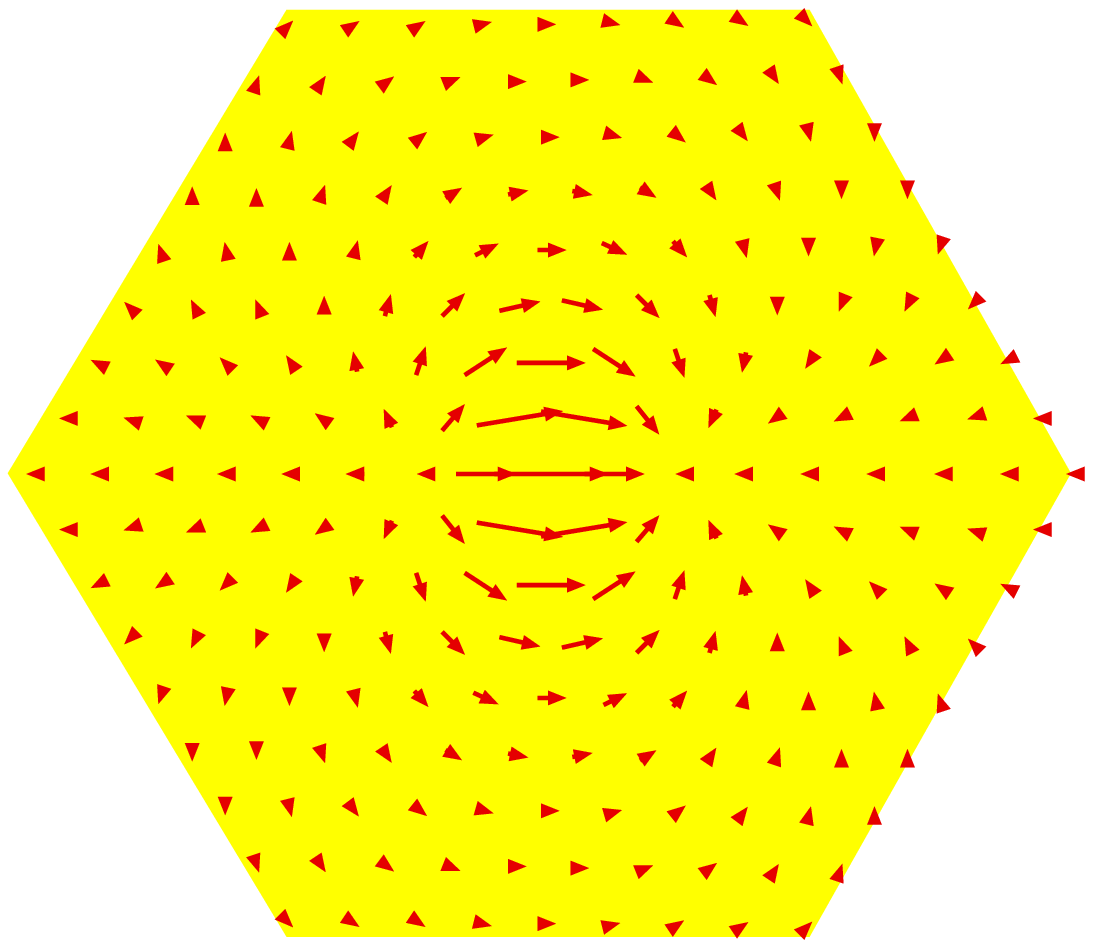}
}
\end{minipage}
\smallskip

\begin{minipage}{0.95\hsize}
\caption[] { \label{patt} \small
Distribution of induced current density (left) 
and corresponding
local electric field (right) in the vortex core at frequency 
$\omega = 0.3 \Delta/\hbar$. The external electric field
$\de \vec{E}_{\omega}(t)=\de\vec{E}_0\cos\omega t$
points into $\hat\vec{x}$-direction.
The top row is the dissipative response
($\sim \cos \omega t$),
and the lower row is the reactive 
response ($\sim \sin \omega t$).
Distances from the center
extend up to $6.3\xi_0$.}
\end{minipage}
\end{figure}
Fig. \ref{patt}
shows typical results for the current and field patterns induced by
the external electric field, 
$\delta{\bf E}_{\omega}(t)=\delta{\bf E}_0\cos{\omega t}$.
The linear response decomposes into an {\em in-phase} 
($\propto \cos{\omega t}$) and {\em out-of-phase} 
($\propto \sin{\omega t}$) response.
The in-phase current response at position ${\bf R}$
determines the local time-averaged energy transfer between
the external field and the electrons, while
the out-of-phase current response is non-dissipative. 
The in-phase and out-of-phase 
currents show characteristically 
different flow patterns.
The non-dissipative current is nearly
homogeneous (lower left),
whereas the pattern of dissipative currents
has qualitatively the form of a
vortex-antivortex pair (upper left), which is consistent with a rigid
shift in $\hat\vec{y}$-direction
of the equilibrium current density. Additionally, this
pattern is strongly deformed at higher frequencies (see Fig. \ref{pattall} below).
The right-hand side of Fig. \ref{patt} shows the total electric field,
which consists of the homogeneous external field and the
internal field due to charge fluctuations induced by the external field.
Note that the induced field is predominantly
out-of-phase and  exceeds the external field in the center of the vortex core.
For higher frequencies, $0.5\Delta \lsim \hbar \omega \lsim 2\Delta $,
the induced dipolar field evolves from
an out-of-phase dipole along the direction of the applied field to an
in-phase dipole opposite to the applied field, thus
reducing the external field by roughly one half.
The induced field vanishes rapidly for
frequencies above the gap edge.
These features of
the current response are robust 
and appear in all our calculations,
independently of the temperature or mean free path.
The charge density fluctuations responsible for the 
induced field are of the order $e\frac{\Delta}{E_f}\delta/\xi_0^2$, 
where $\delta= e\mid\delta\vec{E}_0\mid\xi_0\Delta/(\hbar\omega)^2$
is a small dimensionless para\-meter that measures the amplitude
of the perturbation. Thus, only a fraction $e\frac{\Delta}{E_f}\delta$
of an elementary
charge accumulates periodically in time over an area of $\xi_0^2$,
which is consistent with the condition of
local charge neutrality.\cite{Gorkov75}
Note that the induced charge 
in the vortex core resulting from particle-hole 
asymmetry, as discussed recently,\cite{Khomski95}
is of order $e(\frac{\Delta}{E_f})^2$.
The in-plane current and field response 
(upper row of Fig. \ref{patt})
shows that the local dissipation
can be either positive (``hot spots'' in which the in-phase current
is parallel to the electric field), as well as
negative (``cold spots'' in which the dissipative current
is anti-parallel to the electric field),
and is dominated by dissipation in the vortex center.
The net dissipation is positive, and is obtained by integrating
$\vec{j}\cdot\vec{E}$ over the vortex.
At higher frequencies the dissipation is dominated by higher energy
bound states, and is less localized in the vicinity of the vortex
center.

The electromagnetic response of the vortex-core
is due to an interplay between collective dynamics
of the order parameter and the dynamics of
the Caroli-de Gennes-Matricon bound states.
At frequencies below $0.5\Delta/\hbar $ the order parameter
performs a nearly homogeneous oscillation perpendicular to the
applied field, with a velocity that is $90^o$ out of phase
with the applied field.
To show the deviations of the self-consistent order parameter
from a rigidly oscillating vortex structure, on 
which all prior vortex dynamics calculations are based,
we introduce the displacement vector $\delta \vec{R}_0(t)$
defined by
$\delta\Delta(\vec{R},t)=\delta\vec{R}_0(\vec{R},t)
\cdot\grad\Delta_{0}(\vec{R})$.
The velocity field of the order parameter is then
$\delta \vec{v}(\vec{R},t)=\partial_t \delta \vec{R}_0(\vec{R},t)$.
\begin{figure}[tb]
\centerline{\epsfxsize0.4\hsize\epsffile{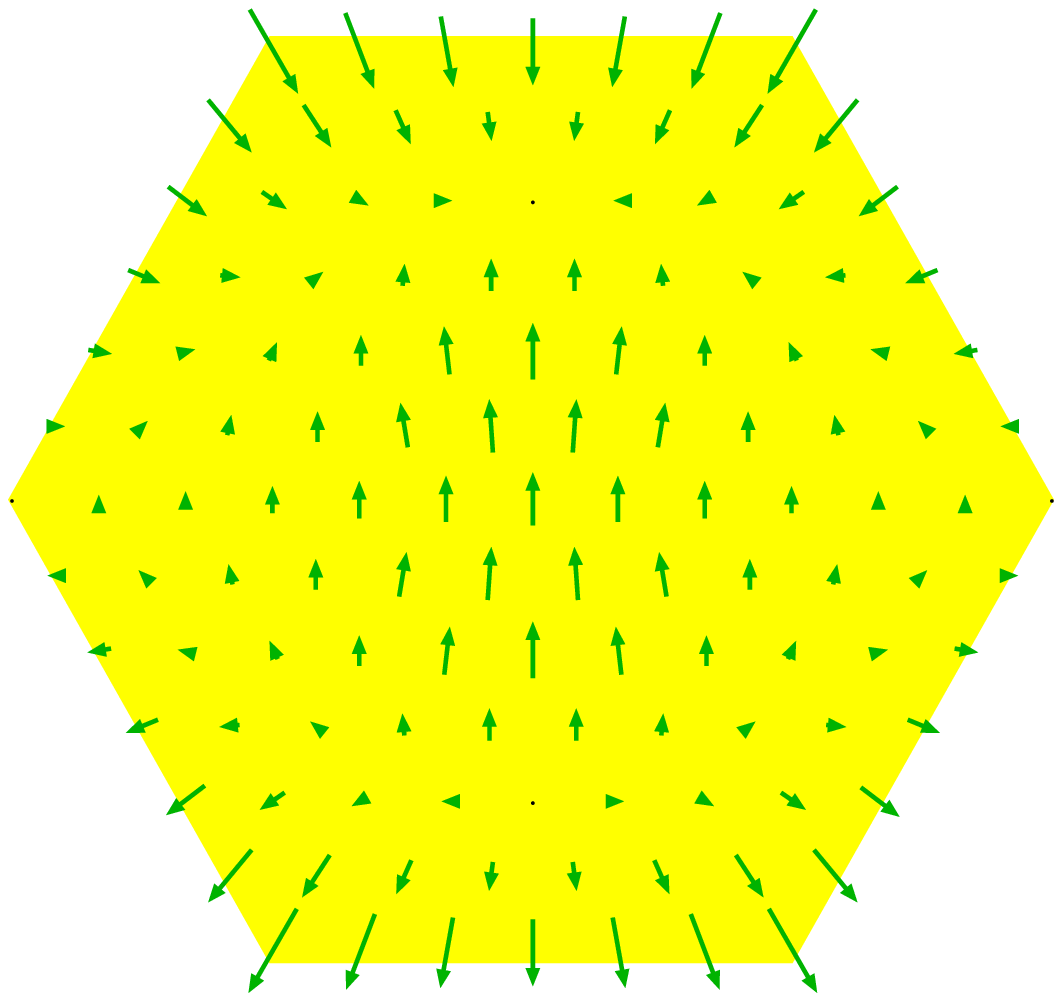}
\hspace{0.1\hsize}
\epsfxsize0.4\hsize\epsffile{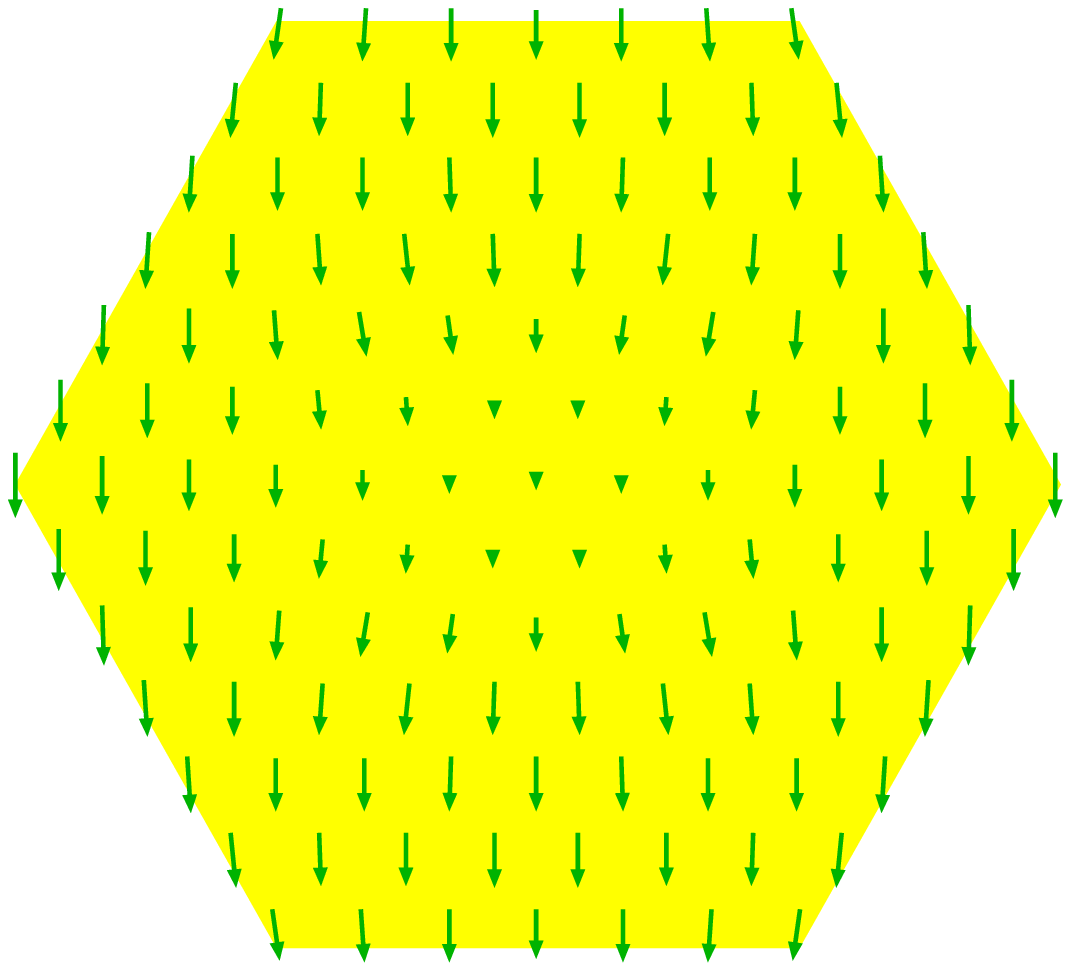}}
\begin{minipage}[t]{0.95\hsize}
\caption[] { \label{ordpar} \small
Velocity field of the order parameter for
$\omega = 1.2 \Delta/\hbar $, showing strong order
parameter deformation.
The left panel shows the in-phase motion of
the velocity field, and the right panel shows
the out-of-phase velocity.
Distances from the center
extend up to $4.7\xi_0$.}
\end{minipage}
\end{figure}
In Fig. \ref{ordpar} we show that the order parameter
velocity field for $\omega =1.2 \Delta/\hbar $
is strongly deformed. The vortex core is not 
rigid, but oscillates in-phase
and perpendicular to the applied field.
The velocity of the vortex center
is of the order of $v_f \delta $,
so the response is far from
that of a stationary vortex.
The amplitude of the vortex
core oscillation increases with
decreasing $\omega$.
However, for $\hbar\omega \gsim \Delta$, both the phase
and amplitude of the vortex core oscillation 
decrease and approach zero above $\hbar\omega\sim 2\Delta $.

\begin{figure}[tb]
\centerline{\epsfxsize0.90\hsize\epsffile{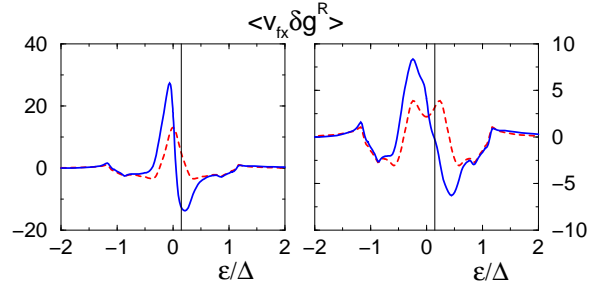}}
\vspace{1mm}
\begin{minipage}{0.95\hsize}
\caption[]{\label{spectra}\small
Spectra for the dissipative part of the 
current density, $\langle v_{fx}\delta g^{R}\rangle$,
at the vortex center (left) and at $0.8\xi_0$ in $\hat\vec{x}$-direction
(right). The vertical lines correspond to $\epsilon=\omega/2$.
The solid curves are the fully self-consistent 
spectral response. The dashed lines assume a frozen
order parameter.}
\end{minipage}
\end{figure}

The coupling between order parameter response and
the bound states is demonstrated in Fig. \ref{spectra},
which compares the nonequilibrium spectral current density
(retarded Green's function, averaged with the weight $v_{fx}$
over the Fermi surface) obtained from a
self-consistent calculation with that
obtained from a non-self-consistent 
calculation that takes into account the
field-induced transitions within the
band of Caroli-de Gennes-Matricon
bound states, but freezes the order
parameter degrees of freedom.
The corresponding contribution to the
current density is obtained by multiplying
these functions by 
$-\frac{1}{2\pi}\tanh\frac{\epsilon-\omega/2}{2T}$ and
integrating over $\epsilon$.\cite{Rainer95}

The dissipative part of the response function
shows peaks in the spectral current density associated
with the band of bound states in the vortex core.
The spectral weight near $\epsilon=\omega/2$
does not contribute significantly to the current response.
Fig. \ref{spectra} shows that the bound state band is shifted
to lower energy by the self-consistent response
of the order parameter, leading to an 
enhanced dissipative current.

\begin{figure}[tb]
\noindent
\begin{minipage}[t]{4.0cm}
\centerline{
\epsfxsize3.9cm
\epsffile{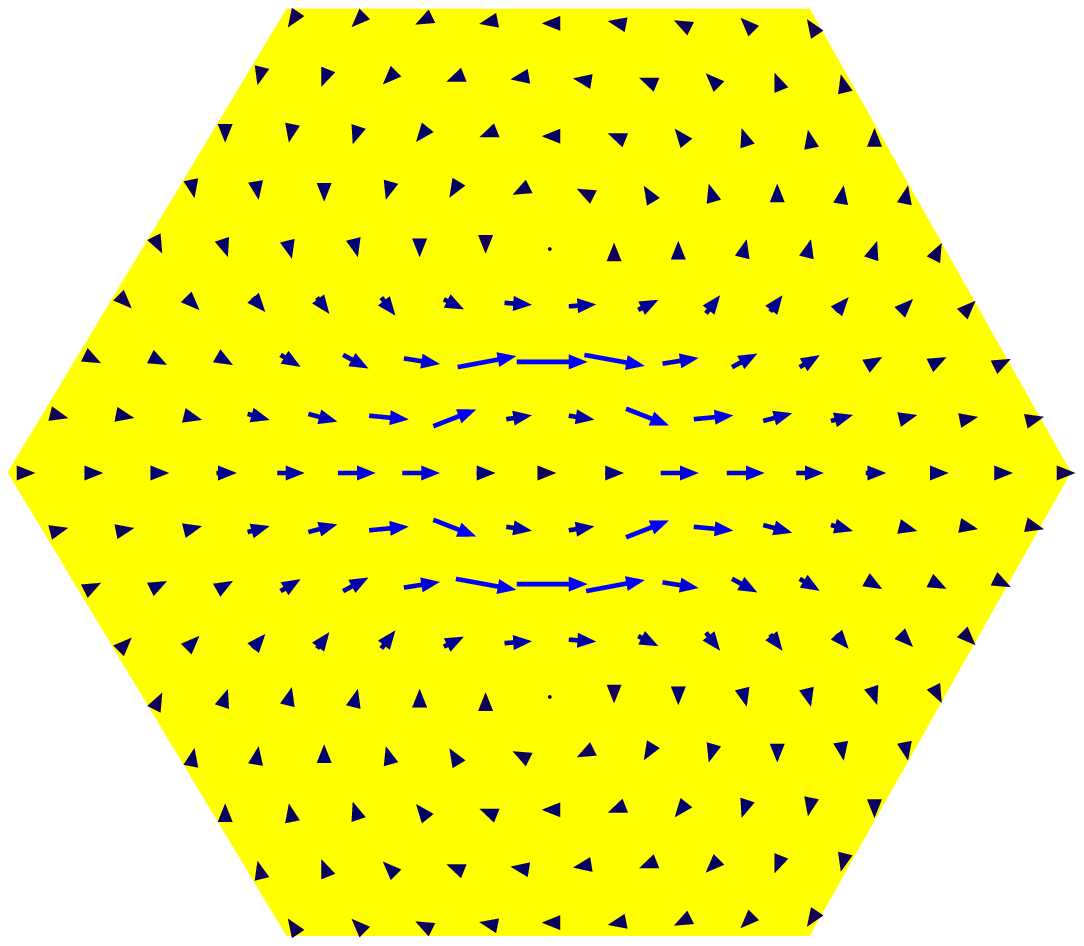}
}
\end{minipage}
\begin{minipage}[t]{4.0cm}
\centerline{
\epsfxsize3.9cm
\epsffile{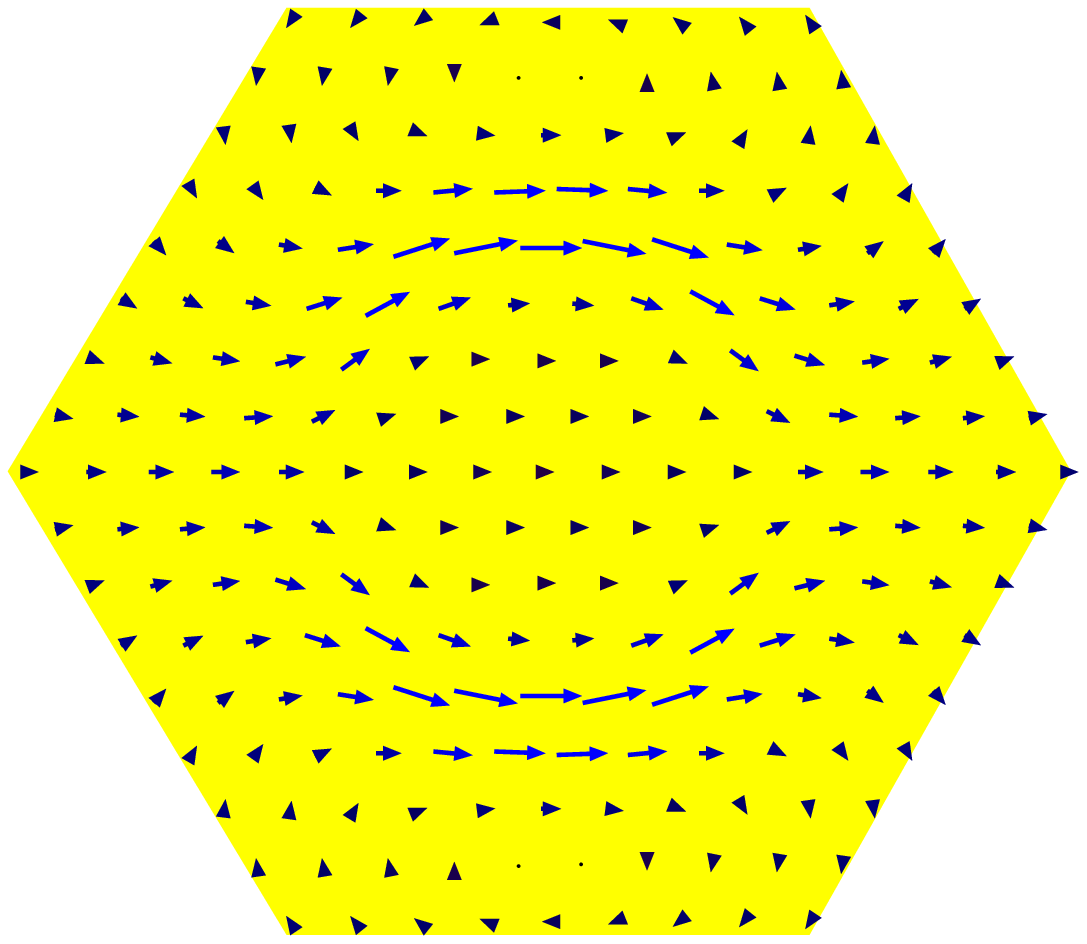}
}
\end{minipage}

\medskip
\begin{minipage}{0.95\hsize}
\caption[] { \label{pattall} \small
Distribution of induced dissipative current density in the vortex core
at frequencies $\omega = 1.1 \Delta/\hbar $ (left) 
and $1.5 \Delta/\hbar $ (right).
The current density at $\omega=1.5 \Delta/\hbar $ is scaled by
$2.5$ relative to the current density at $1.1\Delta/\hbar$.
Higher energy transitions between bound states produce
dissipative currents at larger distances from the vortex
center.
Distances from the center
extend up to $6.3\xi_0$.}
\end{minipage}
\end{figure}

The dissipative currents show nontrivial structure 
for frequencies in the range $\Delta < \hbar \omega <2 \Delta $.
Fig. \ref{pattall} shows that the dissipative currents
flow predominantly in regions where bound states with energy
$\epsilon_{bs}=\hbar \omega/2$ are localized.
We interpret these structures in terms of impurity-mediated
transitions between the Van Hove band edges (shown in Fig. \ref{Nofeps})
of the bound-state bands of the equilibrium vortex.
The transition rate increases with decreasing  
frequency, and for $\hbar\omega$ comparable with
the width of the zero energy bound state the dissipation is
determined by the spectral dynamics of the
zero energy bound states at the vortex center.
The spectral response of these states (Fig. \ref{spectra})
leads to a dramatically
enhanced dissipation near the vortex center,
which is much larger than the normal-state dissipation.

\begin{figure}[tb]
\centerline{\epsfxsize0.90\hsize\epsffile{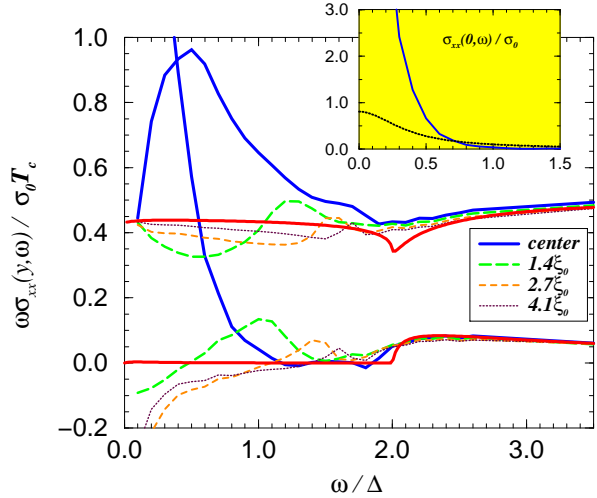}}
\begin{minipage}{0.95\hsize}
\caption[]{\label{sigma}\small
Local conductivity of a vortex as a function of
frequency. The lower set of 
curves shows $\omega\,\mbox{Re}\sigma_{xx}$
as a function of distance from the center (shown in the
legend) along the $\hat{\vec y}$ direction.
The upper set of curves shows $\omega\mbox{Im}\sigma_{xx}$.
The thick red curves
give the real and imaginary parts of the conductivity for
a homogeneous superconductor with the same mean-free-path.
The inset shows a comparison of the Drude absorption (black dotted)
of a normal metal with the enhanced absorption at the center of 
the vortex core. Note that
$\sigma_0=2e^2N_fv_f^2$ and $\ell=10\xi_0$.
}
\end{minipage}
\end{figure}

Fig. \ref{sigma} shows the conductivity in the vicinity of
the vortex core as a function of frequency and distance
from the vortex center. At the vortex center 
the real part of $\sigma_{xx}$ (absorption)
increases dramatically at low frequencies.
Further away from the center the absorption is 
a maximum at a frequency
corresponding to transitions between states near
the Van Hove peaks of the bound state band. 
For comparison we show the conductivity
of a homogeneous superconductor with the same 
mean-free path, which shows an absorption edge
at $\hbar\omega=2\Delta$.
Also note that the low-frequency absorption in the vortex
center is much larger than the Drude
conductivity of the normal state (inset of Fig. \ref{sigma}).
The conductivity sum-rule is obeyed; the apparent
excess weight of $\int\,\mbox{Re} \sigma_{xx}(\omega)d\omega$
is compensated by a negative delta function contribution
associated with counterflowing supercurrents near the vortex
center.
Fig. \ref{sigma} also shows
the enhanced supercurrents in the vortex center, 
compared to those of a homogeneous superconductor,
for frequencies comparable with the width
of the states near zero-energy.

Finally we note that the order parameter response at low frequencies
is mostly transverse to the electric field and that the 
dissipative current
in the core is predominantly parallel to the applied field., i.e. there are no
charge currents in $\hat\vec{y}$-direction related
the order parameter motion.
However, there is a substantial energy flux in the vortex core.
Our calculations show, that the energy flux is predominantly in
direction of the order parameter oscillation and confined in the
vortex core within a few coherence lengths.
The dipolar form of the dissipative current pattern
leads to backflow currents far apart from the vortex core, which
result locally in ``cold spots''.
Thus, the states in the vortex core extract energy from
the external field, and transport
this energy several coherence lengths away from the vortex center
in $\hat{\vec y}$-direction. The net absorption is
determined by inelastic processes at the scale
of the coherence length.

The work of ME and JAS was supported in part by the STC
for Superconductivity through NSF Grant no.\ 91-20000.
DR and JAS also acknowledge support from the
Max-Planck-Gesellschaft and the Alexander von
Humboldt-Stiftung, and ME from the Deutsche Forschungsgemeinschaft.

\vspace*{-0.70cm}

\bibliographystyle{unsrt}

\begin{thebibliography}{10}
\bibitem{Rainer95}
There is also a contribution to the current density
from the ``anomalous'' Green's function which is
not shown here. 
The spectral response shown in Fig. 3 dominates the
dissipative current density in the core, while the
anomalous response contributes primarily to the
reactive response at low frequencies. For more
detail see Ref. \onlinecite{Eschrig97}
\end{thebibliography}

\end{multicols}
\end{document}